# Fabrication of three-dimensional high-aspect-ratio structures by oblique-incidence Talbot lithography

**RYU EZAKI,[1] NAOKI URA,[1] TSUTOMU UENOHARA,[1] YASUHIRO MIZUTANI,[1,*] YOSHIHIKO MAKIURA,[2] AND YASUHIRO TAKAYA[1]**

[1]*Osaka University, Department of Mechanical Engineering, 2-1 Yamada-oka, Suita, Osaka, Japan*
[2]*KURABO INDUSTRIES LIMITED, 14-30 Shimokita, Neyagawa, Osaka, Japan*
[*]*mizutani@mech.eng.osaka-u.ac.jp*

**Abstract:** Developing a suitable production method for three-dimensional periodic nanostructures with high aspect ratios is a subject of growing interest. For mass production, Talbot lithography offers many advantages. However, one disadvantage is that the minimum period of the light intensity distribution is limited by the period of the diffraction grating used. To enhance the aspect ratio of fabricated nanostructures, in the present study we focus on multi-wave interference between diffracted waves created using the Talbot effect. We propose a unique exposure method to generate multi-wave interference between adjacent diffraction orders by controlling the angle of incidence of an ultraviolet (UV) light source. Using finite-difference time-domain simulations, we obtain fringe patterns with a sub-wavelength period using a one-dimensional periodic grating mask. Moreover, we demonstrate the practical application of this approach by using UV lithography to fabricate sub-wavelength periodic structures with an aspect ratio of 30 in millimeter-scale areas, indicating its suitability for mass production.

## 1. Introduction

The structural properties of high-aspect-ratio periodic nanostructures have drawn considerable interest in recent years due to their applicability to various fields such as medical devices [1], genetic engineering [2-4], functional surfaces [5-9], and optical devices [10-13]. In response to this trend, a number of studies have investigated mass production technologies related to highly flexible three-dimensional (3D) periodic nanostructures. Of these, interference lithography, which is a powerful tool for fabricating periodic nanostructures, has attracted significant attention since the 1970s because it uses a simple lithography process that involves periodic light patterns generated via multi-wave interference, and is thus well-suited to fabricating periodic structures [14].

In the simplest version of this technique, two coherent waves interfere with each other to form periodic fringe patterns in space. Then, by exposing the photoresist to this pattern, a one-dimensional (1D) periodic structure that covers a large area can be immediately fabricated. Since this technique was first reported, numerous attempts have been made to employ it in broad applications. For example, the process known as multiple interference pattern exposure, which exposes the photoresist to two or more different patterns, is a well-known method for fabricating complex structures such as nano-needle arrays [15,16]. Similarly, the multiple-wave interference method, in which interference patterns are generated by three or more coherent waves, is another practical application that can be used for fabricating 3D nanostructures [17-19].

However, since previous investigations have shown that interference lithography requires complex optical setups to maintain high coherence, several more recent studies have focused on optical phenomena in the search for ways to generate interference patterns using simple optical setups. These include the near-field holography process, which makes use of 1D periodic gratings to generate fringe patterns with two adjacent order diffracted waves. However, the available period of the grating mask used in this process is limited [20,21]. Other studies have attempted to use two or more grating masks to generate multiple-wave interference patterns with a portion of the diffracted waves, but low light source utilization efficiency remains a problem [22,23].

One alternative, known as the Talbot lithography process, makes use of grating masks to generate multiple-wave interference and is considered well-suited to the fabrication of 3D structures [24]. In this process, which was reported for the first time by Jeon et al. in 2004 [24,25], 3D periodic nanostructures can be fabricated over a large area at one time by a process known as "printing periodic light intensity distribution", which is generated by plane wave transmission via a grating mask [26,27]. However, this Talbot method is disadvantageous in terms of fabrication flexibility because the aspect ratios of fabricated structures are limited by the Talbot effect period. As a result, several exposure methods have been developed to enhance the flexibility of Talbot lithography in spite of its complex lithography process.

One example is the multiple exposure method, which is a useful lithography technique in which two or more different interference patterns are used to fabricate complex periodic structures [28,29]. Another practical application that can be used to fabricate high-resolution periodic patterns is displacement Talbot lithography, which is often used to fabricate protective films in etching processes [30-33]. However, no results have yet been reported on lithography processes that can be used to fabricate periodic nanostructures with high aspect ratios.

In an effort to increase the aspect ratios of structures produced via Talbot lithography, our research focused on the incidence angle flexibility of the ultraviolet (UV) light source based on the notion that by controlling the incidence angle, we could eliminate higher-order diffracted light and generate two-wave interference using diffracted waves with adjacent diffraction orders.

In the sections that follow, this paper describes the incidence angle ranges and grating pitches necessary to generate the two-wave interference needed to fabricate high-aspect ratio structures. First, we report on numerical analyses conducted by controlling the incidence angle in order to theoretically demonstrate the generation of fringe patterns by multiple-wave interference. Then, we describe lithography experiments that were performed to experimentally demonstrate the successful fabrication of high-aspect-ratio periodic structures.

## 2. Oblique incidence Talbot lithography theory

Figure 1 describes the Talbot lithography exposure method used for fabricating high-aspect-ratio periodic structures. In Fig. 1(a) we can see the typical exposure geometry, in which a 1D periodic grating mask and an incident light beam are vertical with respect to the grating surface. In this method, the diffracted waves transmitted by the grating mask generate 2D periodic light distributed in the xz plane and spread it evenly along the y-axis [34,35]. This indicates that the typical Talbot lithography method can be used to fabricate several periodic structure variations simply by changing the grating mask period, even though it is impossible to fabricate high-aspect-ratio periodic structures using wavelength-scale periodic gratings. As a result, the normal Talbot lithography method is unsuitable for fabricating periodic nanostructures with high aspect ratios.

In order to increase the aspect ratio of the fabricated structures, we propose a new exposure method that works by controlling the incident light angle in order to integrate the interference light pattern produced by the Talbot effect. Figures 1(b) and 1(c) show interference patterns with incidence angle variations. In the case of a medium of air and incidence angle of $\theta_i$, the $m$th order diffracted wave angle $\theta_{d,m}$ can be obtained as

$$\theta_{d,m} = \mathrm{Sin}^{-1}\left(\sin\theta_i + m\frac{\lambda}{d}\right), \qquad (1)$$

by the grating equation, where $d$ is the grating pitch and $\lambda$ is the wavelength. This indicates that the diffracted wave angle can be controlled by $\theta_i$. Moreover, for a sufficiently large $\theta_i$, the +1st order diffracted wave disappears, and two-wave interference is generated with the -1st and 0th order diffracted waves. In this case, as shown in Fig. 1(c), we can obtain fringe patterns. In Talbot lithography, the refractive index n of the photoresist differs from that of air, so the angle of the fringe pattern $\theta_r{}'$ becomes

$$\theta_r{}' = \frac{1}{2}\left\{\mathrm{Sin}^{-1}\left(\frac{1}{n}\sin\theta_{d,0}\right) + \mathrm{Sin}^{-1}\left(\frac{1}{n}\sin\theta_{d,-1}\right)\right\}. \quad (2)$$

This indicates that both the structure angle and the aspect ratio of periodic structures can be controlled by the incidence angle.

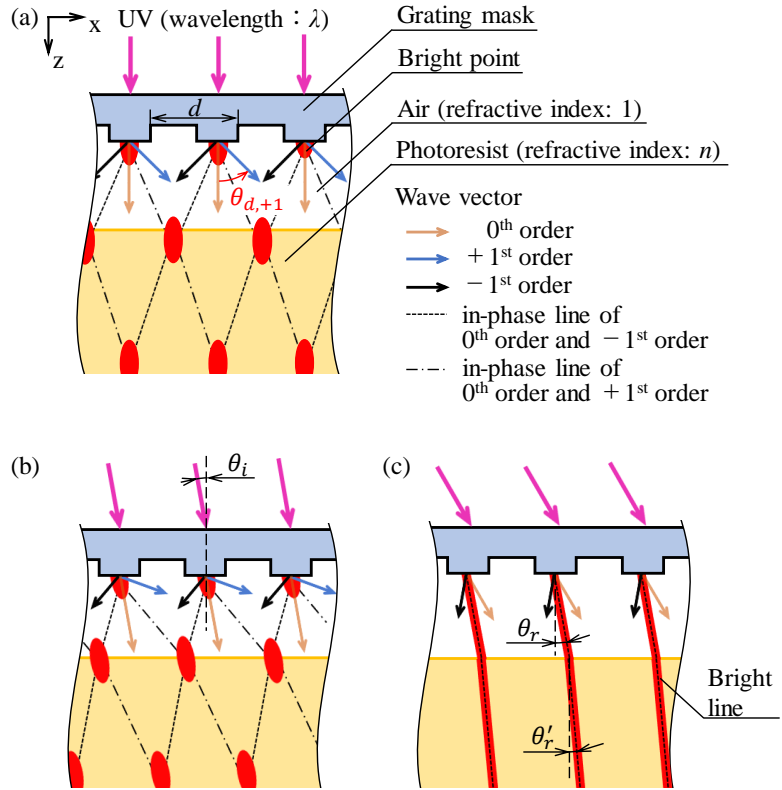

Fig.1. Oblique-incidence exposure method for fabrication of high-aspect-ratio periodic structure. (a) Talbot effect, (b) oblique-incidence Talbot effect with $\theta_{d,+1} < 90°$, and (c) $\theta_{d,+1st} \geq 90°$.

## 3. Analyses conducted to fabricate high-aspect-ratio structures

### 3.1 Analysis of the range ($\theta_i$, $d$) needed to generate fringe patterns via two-wave interference

To analyze the conditions necessary to generate fringe patterns via two-wave interference, a UV laser beam from a diode-pumped solid-state laser (UV-F-360, CV laser, CNI Optoelectronics Tech.) with a wavelength of $\lambda = 360$ nm and a 1D periodic phase mask with a grating pitch of $d = 400$ nm was used to reduce the effect of higher-order diffracted waves. Figure 2 shows the incidence angle dependence of a diffracted wave with a grating pitch of 400 nm, for which the $m$th order diffraction angle follows Eq. (1). From this result, it can be seen that the diffraction angle increases with increasing incidence angle Moreover, for $\theta_i \geq 5.8°$, the +1st order diffracted wave disappears, and two-wave interference is generated between the 0th and -1st order diffracted waves. Additionally, for $\theta_i \geq 53.1°$, a -2nd order diffracted wave is generated, and multiple-wave interference is generated again. This indicates that for d = 400 nm, fringe patterns are generated by two-wave interference over the range of $5.8° \leq \theta_i \leq 53.1°$.

In an effort to generate vertical fringe patterns via two-wave interference, we analyzed the ($\theta_i$, $d$) conditions shown in Fig. 3. In this figure, the black line indicates the border of the two- and three-wave interference, which is calculated from Eq. (1). The hatched area shows two-wave interference generated with -1st and 0th order diffracted waves, while the gray line shows that the fringe pattern is vertical to the z-axis. From these results, it is clear that vertical fringe patterns can be generated over the range of $180 \leq d \leq 540$ nm.

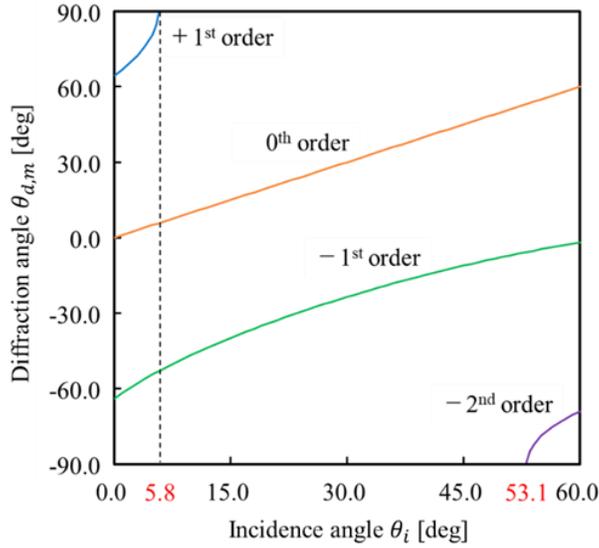

Fig. 2. Range of $\theta_i$ for generating two-wave interference with $d$ = 400 nm.

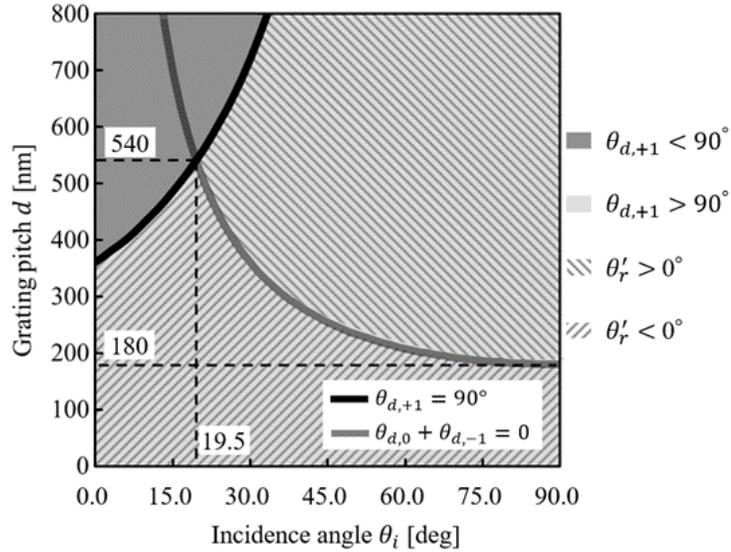

Fig. 3. Range of ($\theta_i$, $d$) for generating horizontal bright lines by two-wave interference.

## 3.2 Verification of incident angle dependence of inclination angle of light intensity distribution

To accurately confirm the generation of vertical fringe patterns, the finite difference time domain (FDTD) method, which uses Maxwell's equations, was used to analyze light intensity distribution. Figure 4 shows a simulation model using a 1D periodic polycarbonate grating mask with a height of 150 nm and a refractive index n of 1.59. Photoresist with an $n$ of 1.72 was set under the grating mask. We also set a perfectly matched layer (PML) as a boundary layer around the simulation area to reduce reflection at the boundary.

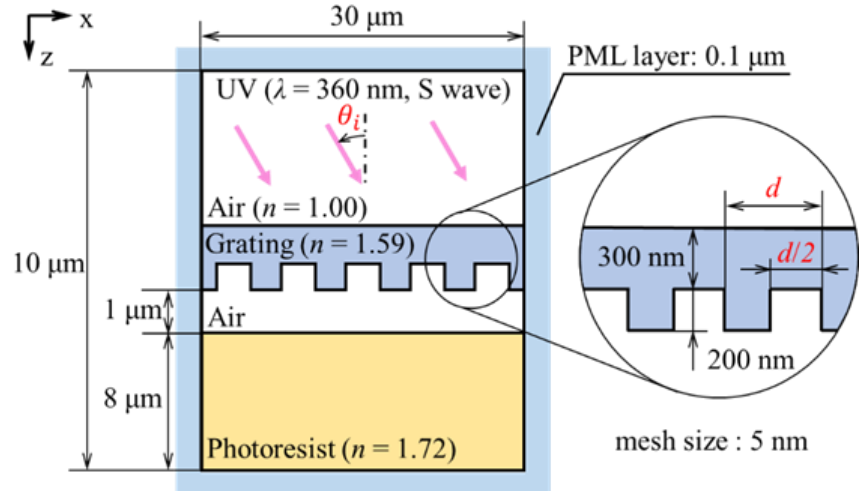

Fig. 4. Simulation model for determining the incidence angle dependence of the aspect ratio of the structure. In order to remove the influence of the non-interference region near the PML layer, only the light intensity distribution within the region $13.5 \leq x \leq 16.5$ µm was considered.

Since the simulation model has a width of 30 µm along the x-axis, the light intensity distribution due to the Talbot effect is only useful in the region of $13.5 \leq x \leq 16.5$ µm, and regions near the right and left sides are not considered. This is because it is necessary to remove the influence of the non-interfering region near the edges. To verify the incidence angle range needed to generate fringe patterns by two-wave interference, the incident angles of the 360 nm wavelength UV laser was set as 0°, 5°, 10°, 20°, 30°, 40°, 50°, and 60°.

Table 1 shows the light intensity distribution simulation results. When $\theta_i = 0$ and 5°, 2D periodic patterns are generated by multiple-wave interference. Moreover, when $10 \leq \theta_i \leq 50°$, fringe patterns are generated by two-wave interference. In contrast, when $\theta_i = 60°$, 2D periodic patterns are generated again. This result is attributed to the occurrence of a -2nd order diffracted wave. Figure 5 shows the incidence angle dependency of the $\theta_r'$ fringe pattern inclination angle. The curved line in the figure shows the theoretical relationship between the incidence angle and the inclination angle of the fringe pattern, which is calculated by Eq. (2), while the marker shows the FDTD simulation results for the inclination angle, which is in good agreement with the theoretical curve. From these results, we determined that fringe patterns could be generated by the oblique exposure method and that fringe pattern inclination angles vary with the incidence angle.

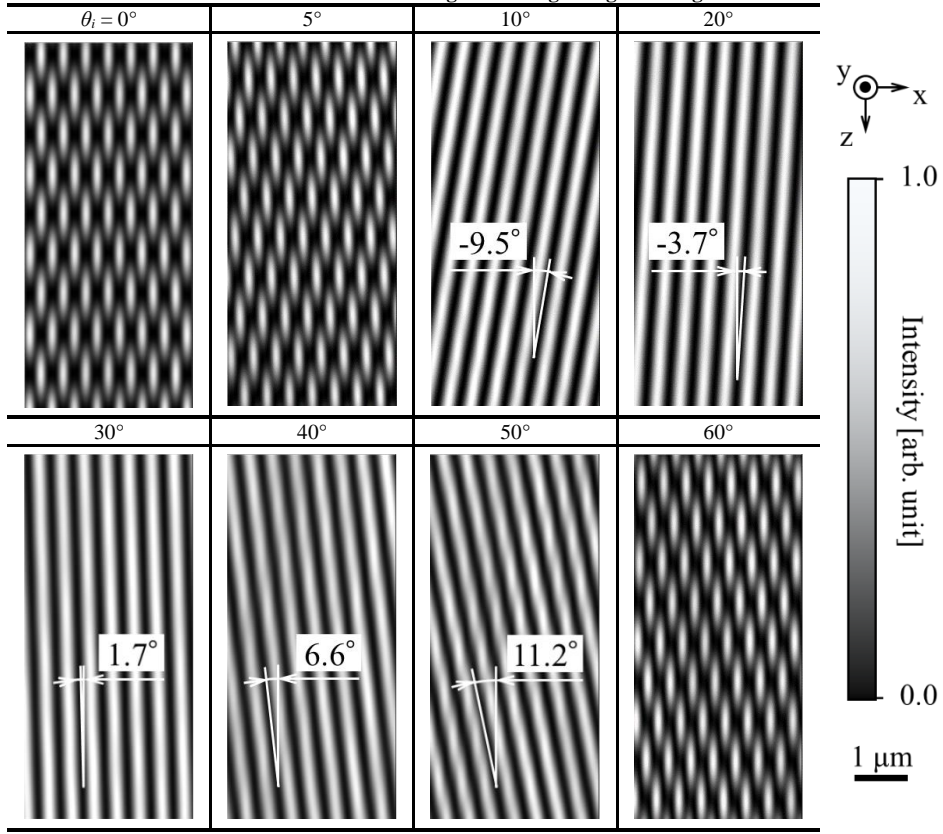

Table 1. Simulation results for determining the $\theta_i$ range for generating two-wave interference.

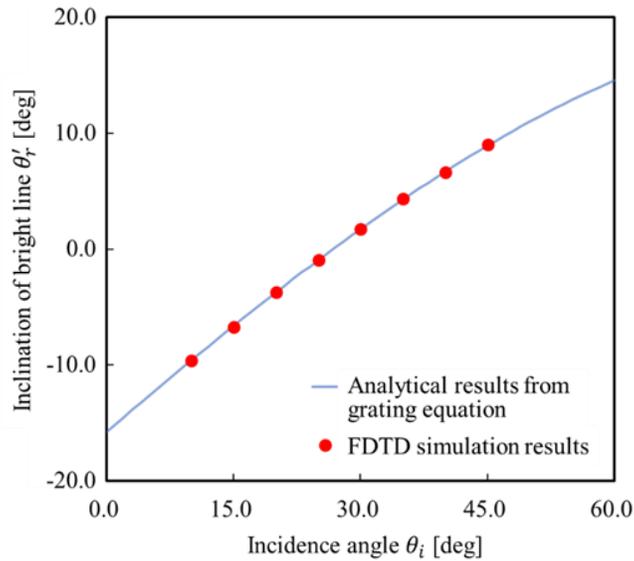

Fig. 5. Comparison of analytical and FDTD results for inclination of bright line. The results indicate that the inclination can be controlled by the incidence angle.

## 3.3 Determination of the grating pitch range required to generate vertical fringes patterns

As explained in the previous section, the FDTD simulation results show the incidence angle dependence of the fringe pattern inclination angle, which follows Eq. (2). Therefore, we used this simulation model to determine the range required to generate vertical fringe patterns via two-wave interference. To generate vertical fringe patterns, we set $(\theta_i, d)$ = (64°, 200), (37°, 300), (27°, 400), (21°, 500) and (17°, 600), which satisfy $\theta_r' = 0°$.

Table 2 shows the light intensity distribution simulation results. When $d$ = 200-500 nm, vertical fringe patterns are generated by two-wave interference. In contrast, when $d$ = 600 nm, the light intensity distribution becomes discontinuous along the z-axis due to the occurrence of a +1st order diffracted wave. From these results, corresponding to the theoretical range of $180 \leq d \leq 540$ nm, we determined that a vertical structure with a high aspect ratio could be fabricated by controlling the incidence angle and grating pitch.

**Table 2. Simulation results of light intensity in photoresist layer to generate vertical bright line by two wave interference.**

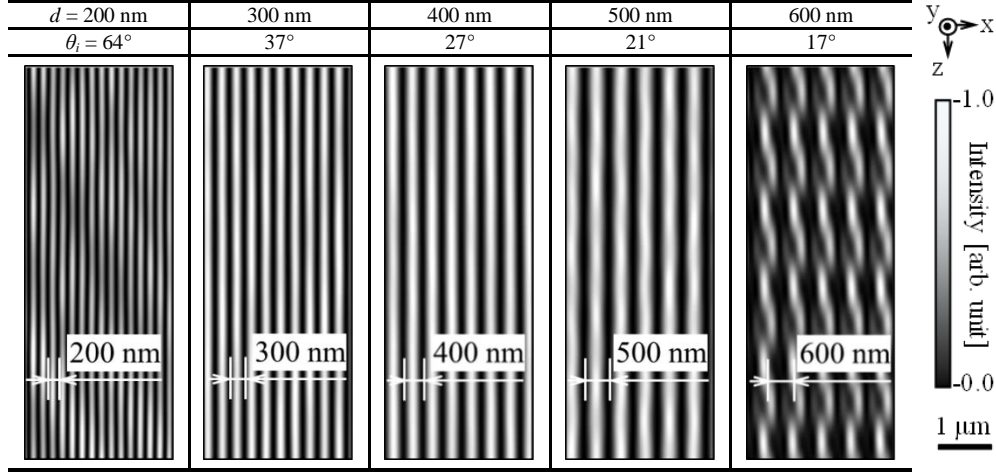

## 4. Fabrication of high-aspect-ratio periodic structures

### 4.1 Verification of incidence angle dependence of aspect ratio of fabricated structures

In this research, lithography experiments were conducted to experimentally demonstrate the successful fabrication of high-aspect-ratio periodic structures. To analyze the effect of higher-order diffracted waves on such fabricated structures, we used a polycarbonate grating mask with a pitch of 750 nm and a blaze height of 150 nm. Under those experimental conditions, which are based on Eq. (1), the +1st order diffracted wave disappears for $\theta_i \geq 31°$ and a -2nd order diffracted wave is generated for $0 \leq \theta_i \leq 90°$. Therefore, the experimental results indicate that the period of the fabricated structure is affected by the -2nd order diffracted wave.

Figure 6 shows the Talbot lithography process used in our experiments. First, as shown in Fig. 6(a), a 6 μm thick coat of positive photoresist (SIPR-3251N-6.0, Shin-Etsu Chemical) was spread on the glass substrate (C218181, MATSUNAMI) by spin coating. Next, as shown in Fig. 6(b), the photoresist solvent was evaporated by prebaking on a hotplate (ACT-200DII, Active). Then, as shown in Fig. 6(c), the obliquely incident UV laser beam from a diode-pumped solid-state laser (UV-F-360, CV laser, CNI Optoelectronics Tech.) with a wavelength of $\lambda$ = 360 nm was masked by a grating, and the light intensity pattern produced by the Talbot effect was printed in the photoresist layer. Note that the total exposure dose was 250 mJ/cm², and the rotating stage (KSP-606M, SIGMAKOKI) positioned under the sample was used to control the incidence angle. Finally, as shown in Figs. 6(d) and 6(e), after developing and rinsing with 2.38% tetramethylammonium hydroxide (TMAH; TAMA CHEMICALS) and distilled water (KISHIDA CHEMICAL), respectively, periodic structures were fabricated in the photoresist layer.

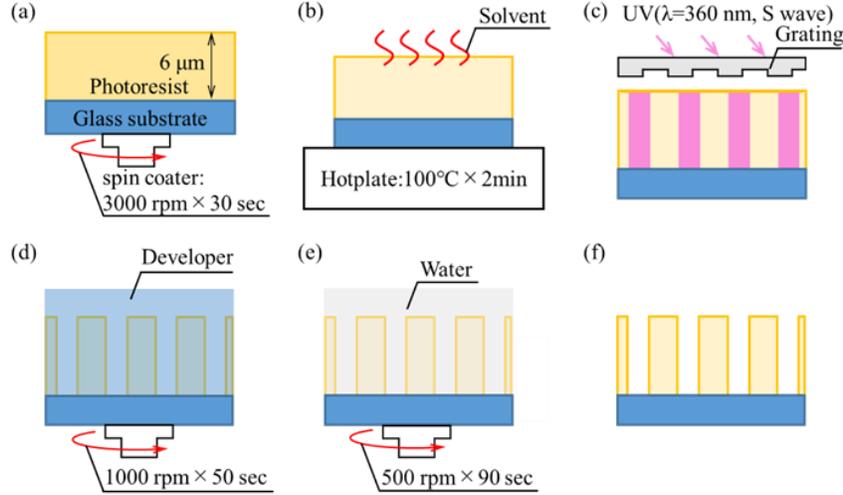

Fig. 6. Lithography process for high-aspect-ratio periodic structure.

Figure 7 shows an overview of the structure fabricated when $\theta_i = 0°$. Due to the reflected light pattern produced by the diffraction phenomenon, we can see that a periodic structure was fabricated over the entire exposure area. Table 3 shows cross-sectional scanning electron microscopy (SEM; JSM-6010PLUS, JEOL) images of fabricated structures produced with incidence angles of 0°, 10°, 20°, 30°, and 40°. From these results, it can be seen that periodic structures were fabricated in both samples and that the inclination angle of the fabricated structures depended on the incidence angle.

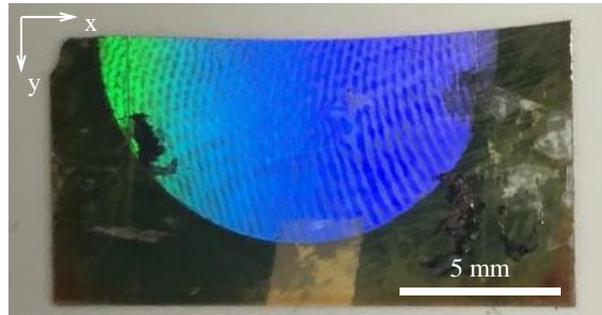

Fig. 7. Overall picture of fabricated structure with $\theta_i = 0°$.

For $\theta_i = 20°$, 30°, and 40°, it is particularly noteworthy that high-aspect-ratio periodic structures were fabricated with heights larger than 6 μm and aspect ratios of about 15. Nevertheless, the -1st order diffracted wave disappears when $\theta_i \geq 31°$. This result can be explained by the interpolation of periodic light patterns. A 2D periodic structure was fabricated for $\theta_i = 20°$, which is consistent with the theory of multiple-wave interference regarding the 0th and ±1st order diffracted waves. On the other hand, a 2D periodic structure was also fabricated for $\theta_i = 40°$, which is believed to be caused by multiple-wave interference between the 0th -1st and -2nd order diffracted waves. These experimental results show that the aspect ratio of the fabricated structure can be controlled by the incidence angle.

Table 3. Cross sections of fabricated structures with *d* = 750 nm.

| $\theta_i = 0°$ | 10° | 20° | 30° | 40° |
|---|---|---|---|---|
| 0.40 μm, 1.92 μm | 0.39 μm, 2.12 μm | 6.90 μm, 0.46 μm | 6.92 μm, 0.41 μm | 6.10 μm, 0.32 μm |

*4.2 Fabrication of vertical structure with high aspect ratios*

To analyze the differences between two-wave and multiple-wave interference, grating masks with *d* = 320 and 750 nm were used to fabricate vertical structures. In these experiments, the lithography process shown in Fig. 6 was conducted, and the experimental setup was as shown in Section 4.1. To fabricate vertical structures, the experimental conditions were set as ($\theta_i$, *d*) = (34°, 320) and (13°, 750) in the exposure process, based on Eq. (2). In this case, fringe patterns produced by two-wave interference were generated for *d* = 320 nm, and 2D periodic patterns produced by multiple-wave interference were generated for *d* = 750 nm.

Figure 8 shows cross-sectional SEM images of the fabricated structures. The vertical structures were fabricated with x-axis periods that were equal to the grating pitch. Moreover, as shown in Fig. 8(a), periodic structures with a linewidth of 170 nm and aspect ratios of about 30 were fabricated for *d* = 320 nm. On the other hand, as shown in Fig. 8(b), due to the existence of the -2nd order diffracted wave, 2D periodic structures were fabricated with a maximum linewidth of 300 nm and an aspect ratio of 20. Based on these results, it can be concluded that the proposed exposure method is a powerful tool for fabricating vertical structures with high aspect ratios, which has also been demonstrated by our previous experimental results.

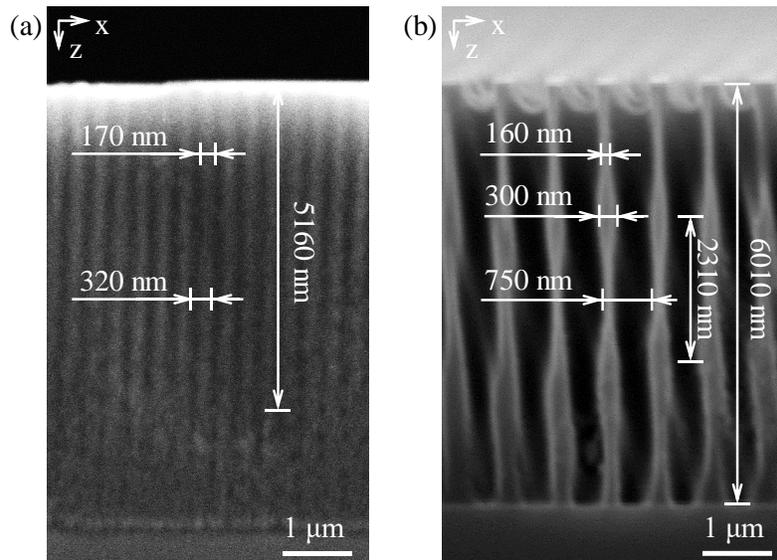

Fig. 8. Cross-sectional SEM images of vertical structures with (a) *d* = 320 nm and (b) *d* = 750 nm.

## 5. Conclusions

Herein, we proposed an oblique incidence exposure method that can be used to fabricate high-aspect-ratio structures. Our FDTD simulation results show that fringe patterns can be generated, and their inclination angles can be controlled by the incidence angle. Moreover, the experimental results showed that 2D periodic structures with different inclination angles could be fabricated in millimeter-scale areas. In particular, it was noteworthy that vertical structures with aspect ratios exceeding 20 could be fabricated with a linewidth of 170 nm. In conclusion, we believe our proposed Talbot lithography method is suitable for use in mass production processes involving 3D periodic structures with high aspect ratios and narrow linewidths.

## Acknowledgments

This work was supported by a Grant-in-Aid from KURABO INDUSTRIES LTD. This work was also supported by JSPS KAKENHI Grant Numbers 18K18811 and 19H02154.

## Disclosures

The authors declare no conflicts of interest.